\documentstyle[11pt,aasms4]{article}

\begin{document}

\def\sec{$^{\prime\prime}$}
\def\min{$^{\prime}$}

\hyphenation{a-ni-so-tro-pic flux---ca-li-bra-ted}

\title{The Spectral Energy Distribution of Normal, Starburst and Active Galaxies}

\author{Henrique R. Schmitt \altaffilmark{1,2,3},
Anne L. Kinney\altaffilmark{2,4},
Daniela Calzetti\altaffilmark{2},
and Thaisa Storchi-Bergmann\altaffilmark{1}}

\altaffiltext{1}{Departamento de Astronomia, IF-UFRGS, CP 15051, CEP91501-970,
Porto Alegre, RS, Brazil}
\altaffiltext{2}{Space Telescope Science Institute, 3700 San Martin Drive,
Baltimore, MD21218} 
\altaffiltext{3}{CNPq Fellow} 
\altaffiltext{4}{Department of Physics and Astronomy, Johns Hopkins University,
Baltimore, MD21218}

\begin {abstract}

We present the results of an extensive literature search of
multiwavelength data for a sample of 59 galaxies, consisting of 26
Starbursts, 15 Seyfert 2's, 5 LINER's, 6 normal spirals and 7 normal
elliptical galaxies. The data include soft X-ray fluxes, ultraviolet
and optical spectra, near, mid$/$far infrared photometry and radio
measurements, selected to match as closely as possible the IUE aperture
(10\arcsec$\times$20\arcsec). The galaxies are separated into 6 groups with
similar characteristics, namely, Ellipticals, Spirals, LINER's, Seyfert
2's, Starbursts of Low and High reddening, for which we create average
spectral energy distributions (SED).

The individual groups SED's are normalized to the
$\lambda$7000\AA\ flux and compared, looking for similarities and
differences among them. We find that the SED's of Normal Spirals and
Ellipticals are very similar over the entire energy range, and fainter
than those of all other groups.  LINER's SED's are similar to those of
Seyfert 2's and Starbursts only in the visual to near-IR waveband,
being fainter in the remaining wavebands. Seyfert 2's are similar to
Starbursts in the radio to near-IR waveband, fainter in the visual to
ultraviolet, but stronger in the X-rays. Low and High reddening
Starbursts are similar along the entire SED, differing in the
ultraviolet, where Low reddening Starbursts are stronger, and in the
mid$/$far IR where they are fainter.

We have also collected multiwavelength data for 4 HII regions, a
thermal supernova remnant, and a non-thermal supernova remnant (SNR),
which are compared with the Starburst SED's.  The HII regions and
thermal SNR's have similar SED's, differing only in the X-ray and far
infrared. The non-thermal SNR SED is a flat continuum, different from
all the other SED's.  Comparing the SED's of Starbursts and HII regions
we find that they are similar in the mid$/$far IR parts of the
spectrum, but HII regions are fainter in the radio and X-rays.
Starbursts are also stronger than HII regions in the visual and near-IR
parts of the spectrum, due to the contribution from old stars to
Starbursts.

The bolometric fluxes of the different types of galaxies are calculated
integrating their SED's. These values are compared with individual
waveband flux densities, in order to determine the wavebands which
contribute most to the bolometric flux. In Seyfert 2's, LINER's and
Starbursts, the mid$/$far IR emission are the most important
contributers to the bolometric flux, while in normal Spirals and
Ellipticals this flux is dominated by the near-IR and visual
wavebands.  Linear regressions were performed between the bolometric
and individual band fluxes for each kind of galaxy. These fits can be
used ine the calculation of the bolometric flux for other objects of
similar activity type, but with reduced waveband information.

\end{abstract}


\section{Introduction}

With the present availability of large databases, including satellite
observations at wavebands that cannot be observed from the ground, like
X-rays, ultraviolet, mid$/$far IR, it is possible to construct spectral
energy distributions (SED's) of galaxies over 10 decades of
frequency.  The study of the continuum emission of galaxies over such a
broad range of frequencies is important for a good determination of the
bolometric luminosity of these objects. Also, the SED's can be used to
study the energy output at different wavebands, as well as a means to
distinguish galaxies of different activity classes.

Previous works, like those of Edelson \& Malkan (1986) and Sanders et
al. (1989) investigated the SED's of AGN's. Edelson \& Malkan (1986)
analyzed a small group of Seyfert 1's, Seyfert 2's and Quasars, but did
not include radio and X-ray fluxes in their SED's, while Sanders et al.
(1989) presented radio to X-ray SED's for a sample of Radio Loud and
Radio Quiet Quasars.

While the SED's of high luminosity AGN's have been relatively well
studied, little has been done on the SED's of Starbursts, Seyfert 2's,
LINER's and Normal galaxies.  Mas-Hesse et al. (1994,1995) have
presented a radio to X-ray multiwavelength analysis of Seyfert 1's,
Seyfert 2's, Starbursts and Quasars, but with relatively sparse data
points to cover the frequency range. They found that these objects can
be divided into two major groups, those objects where the far infrared
emission dominates the SED (Seyfert 2's and Starbursts), and those
objects where the UV and X-ray have fluxes comparable to the far
infrared (QSO's and Seyfert 1's). They also point out that Seyfert 2's
and Starbursts have similar SED's, but Seyfert 2's are brighter in the
X-rays.

Another multiwavelength analysis of Starbursts, Seyferts, LINER's,
Quasars and normal galaxies was made by Spinoglio et al. (1995). They
do not use radio and X-ray fluxes and only include a small number of
wavebands. Spinoglio et al. (1995) also apply a correction to include
the flux of the entire galaxy in their analysis, which is uncertain.
Their results show that the nonstellar radiation at 2-3$\mu$m
correlates with the IRAS colors, which produces a sequence of colors,
that runs from normal galaxies to Seyfert 2's, Seyfert 1's and Quasars.
Starbursts fall outside this sequence, because they have an excess of
60$\mu$m emission. In contrast to Mas-Hesse et al. (1995), Spinoglio et
al. (1995) found that in the mid$/$far infrared Seyfert 2's are more
similar to Seyfert 1's than to the Starbursts, which they attribute to
the fact that Seyferts are heated by a single source, while Starbursts
have an extended heating region.

In this paper we present the Spectral Energy Distribution from radio
($\nu\approx10^8$ Hz) to soft X-rays ($\nu\approx10^{18}$ Hz) of a
sample of galaxies including Starbursts, Seyfert 2's, LINER's, normal
Spirals and Ellipticals.  While the data were selected in order to
match as closely as possible the IUE aperture
(10\arcsec$\times$20\arcsec), the match is indeed not very good, and is
the main challenge in assembling and interpreting such a data set.  The
galaxies were divided in 6 groups according to activity class and, in
the case of quiescent galaxies, according to morphology, for which we
create average SED's.  These average SED's are compared
to verify whether we can
use the SED's to separate different activity classes. We also present
the SED's of HII regions, a thermal SNR and a non-thermal SNR, which
are compared to Starburst SED's.

In Section 2 we describe our sample and in Section 3 we discuss the
data and aperture effects. The SED's of the individual groups are
described in Section 4 and compared in Section 5. In Section 6 we
describe the HII regions and Supernova Remnants SED's and compare them
with Starbursts SED's.  A statistical comparison between the SED's of
galaxies with different activity classes is given in Section 7. The
bolometric luminosities are discussed in Section 8, while in Section 9
we give the summary.

\section{The Sample}

The galaxies were selected from the catalog of ultraviolet IUE spectra
of Kinney et al. (1993) and from Kinney et al. (1996). We include only
those objects for which we have ground based spectra, observed with
apertures matching that of IUE (Storchi-Bergmann, Kinney \& Challis
1995; McQuade, Calzetti \& Kinney 1995; Kinney et al. 1996).

The sample is composed of 59 objects, with 26 star-forming
galaxies, 15 Seyfert 2's, 5 LINER's, 6 normal spirals, 6 normal
ellipticals and 1 bulge of a spiral (NGC224, which is treated as an
elliptical). Their names, morphological types, activity classes, radii
and velocities relative to the local group of galaxies are given in
Table~1. For objects with composite activity class we
assume that the first class listed in the reference is dominant.

\section{The Data and Aperture Effects}

We searched the literature for X-ray, infrared and radio data of the
sample galaxies, selecting, when possible, data observed with apertures
close to that of the IUE satellite (10\arcsec$\times$20\arcsec).  Note
that although the apertures don't match very well, a comparison between
bolometric fluxes and galaxy diameters (Section 8), shows that these
quantities are relatively independent. Thus the aperture effects do not 
generally dominate
the data.

The UV and optical data (14.5$\leq$ Log $\nu \leq$ 15.5) were obtained
from Table~4 of McQuade et al. (1995) and Table~4 of Storchi-Bergmann
et al. (1995). The UV is composed of IUE spectra in the wavelength
range 1100--3200\AA, while the optical comes from ground based spectra
in the range 3200--10000\AA\, observed with matched apertures. Details
of the observations and reductions are given in the above papers.
Notice that instead of using the spectra, we use only the continuum
fluxes measured on selected points, because we are interested only on
the continuum energy distribution  and not on individual spectral
features. We give in Table~2 the UV and optical continuum fluxes for 8
galaxies of the sample, also observed by the authors, but whose data
was not previously published.

In Table 3 we show the radio data (Log $\nu <$10), available in the
literature, for the objects in Table 1. Since the galaxies were not
observed exactly at the same wavelengths we indicate in the table
header the approximate wavelengths. The information for each entry is
divided in three lines; on the first line we give the flux (in units of
mJy), on the second line we give the actual frequency of the
observation (in units of GHz), and on the third line the aperture
through which it was observed.  On the last column we give the
references from which each entry of the table was obtained, ordered
from left to right, according to the numbers listed in Table~5.  The
apertures for the radio data vary from 3\arcsec\ to apertures of the
order of arcminutes, containing the entire galaxy. This spread in
apertures introduces a large spread in the fluxes, with the smaller
apertures including only nuclear emission and the larger apertures
including also extended radio emission and emission from HII regions
and SNR's along the galaxy disk, significantly increasing the flux.

Millimeter, near infrared (1.2$\mu$m-20$\mu$m) and X-ray data are shown
in Table~4, in the same format as in Table~3. Millimeter data are rare,
being available only for 3 Seyfert~2 galaxies.  These data can be
considered only as additional information for these galaxies, since we
cannot compare them with the other classes of objects. Data in the
near-infrared range (13.5$\leq$ Log $\nu \leq$14.5) are available for
the majority of the galaxies in our sample, usually with apertures very
close to that of IUE.

For the X-ray waveband (Log$\nu>$15.5) we use data from the Einstein
catalog of Fabbiano, Kim \& Trinchieri (1992). We chose to use Einstein
instead of ROSAT data, because it has observations available for a
larger number of galaxies, including almost all the galaxies observed
with ROSAT (the only exception is NGC3256, for which only ROSAT data
are available).  The aperture in the X-ray is problematic, because it
includes the entire galaxy, with both extended emission and sources in
the galaxy disk, farther than the 10\arcsec $\times$20\arcsec\ central
region.  In the case of Starbursts and Seyfert 2's, where most of the
X-ray flux comes from the nuclear region, the aperture does not affect
the results considerably. However, for LINER's and Normal galaxies,
this assumption is not valid, and their X-ray fluxes may be strongly
contaminated by HII regions, Supernova Remnants and X-ray binaries
along the disk of the galaxy.

The X-ray fluxes given in Table 4 are integrated over the entire
waveband (0.2-4.0Kev for Einstein or 0.1-2.4 Kev for ROSAT). In order
to put these fluxes in the same units as the other wavebands, we
assume the X-ray spectrum to be $\propto\nu^{-1}$, and calculate the
flux to the central energy of the band (2.1 kev for Einstein and 1.25
kev for ROSAT). The assumption of a slope of --1 ($\nu^{-1}$)
would underestimate the central energy flux by 40\% if the true slope was
--0.5, or overestimate it by 40\% if the true slope was --1.5.

The IRAS data (12.5$\leq$ Log$\nu \leq$ 13.5), in the mid$/$far IR (12,
25, 60 and 100$\mu$m), were obtained from NED (NASA Extragalactic
Database).  Due to the large aperture through which they were obtained,
which varies from 0.75\arcmin$\times$4.5\arcmin\ at the 12$\mu$m band
to 3\arcmin$\times$5\arcmin\ at 100$\mu$m, these data are challenging
for our analysis. The aperture discrepancy is probably the least
problematic for Starbursts where, the light is concentrated towards the
nucleus according to Calzetti et al. (1995).  Likewise, IR emission
from the Seyfert 2 galaxies is probably also dominated by nuclear
emission. However, the emission from the LINER's and Normal galaxies is
likely strongly contaminated by sources throughout the galaxy disk.

\section{Spectral Energy Distributions}

The sample is divided in six groups: normal Ellipticals, normal
Spirals, Seyfert~2's, LINER's, and high and low reddening Starbursts.
The division between low and high reddening Starbursts is made at
E(B--V)$=$0.4, assuming the values given by Calzetti, Kinney \&
Storchi-Bergmann (1994).  The low reddening group is composed of the
galaxies:  HARO15, MRK357, MRK542, MRK66, NGC1140, NGC3049, NGC5236,
NGC5253, NGC6052, NGC7250 and UGC9560; while the high reddening one is
composed of:  IC1586, IC214, NGC1097, NGC1313, NGC1672, NGC3256,
NGC4385, NGC5860, NGC5996, NGC6090, NGC6217, NGC7552, NGC7673, NGC7714
and NGC7793.

The foreground Galactic extinction for the galaxies in our sample is
small, and no correction is applied. Also, due to the small redshift of
the galaxies, only the data in the wavelength range 1100--10000\AA,
corresponding to the IUE and ground based spectra, were redshift
corrected. No redshift correction was applied to the broad band data.
The possible errors introduced by these factors are minimal and will
not affect the overall analysis.

The individual SED's of normal Elliptical and Spiral galaxies are shown
in Figure~1a. SED's of Seyfert~2's and LINER's are shown in Figure~1b
and those of the Starbursts of low and high reddening are shown in
Figure 1c. From now on we will refer to the low and high reddening
Starbursts as SBL and SBH, respectively.  The SED's are shifted in the
figures by arbitrary constants, for clarity.  The radio and X-ray upper
limits are shown as filled dots. Notice also that we draw a straight
line from the radio to the far-IR 100$\mu$m wavebands. This assumption
do not represent the real SED in the millimeter region, which according
to Antonucci, Barvainis \& Alloin (1990), present a dip around 1 mm.

The SED's were normalized to the flux at $\lambda$7000\AA, which
corresponds to a normalization to the old stellar population
contribution, and are shown in Figure~2.  The average SED's, obtained
from the latter, are shown in Figure~3 and their values are given in
Table~6.  Since the upper limits presented values similar to the real
detections in the same wavebands, we decided to include them in the
averages.

As we can see in the above Figures, the Elliptical galaxies have
similar SED's in the UV to near-IR range, presenting an old, red
stellar population and the UV turn-up. However, in the mid$/$far IR and
radio wavebands there is a large difference between individual SED's.
The differences in the mid$/$far IR can be attributed to different
amounts of dust (Goudfrooij \& de Jong 1995), while in the radio, the
existence of a radio loud nucleus can influence the SED radio tail
significantly. These differences could also be due to the different
apertures through which the observations were taken.  The X-ray fluxes
have some spread, which is due to the large aperture through which they
were observed, including the contribution from sources like X-ray
binaries and the hot gaseous halo (Fabbiano 1989), which extend for
much more than 10\arcsec$\times$20\arcsec.

Normal Spiral galaxy SED's, when compared with the SED's of Elliptical
galaxies, have a considerable spread in the UV to near-IR, which is due
to the presence of HII regions in the disk, close to the nucleus of
some of these galaxies. The X-ray data are available for only two
objects, showing vastly different values of slope from optical to X-ray
and so will not be used in the rest of the analysis.  The mid$/$far IR
emission, like that emission in the Ellipticals, have a large spread,
which can be attributed to both aperture and dust effects. In the radio
waveband, the SED's are very similar, with the exception of NGC598,
which is the higher radio emitter.  This galaxy has a radius far
greater than the other spirals in the sample, implying that the
difference is due to aperture effects.

The Seyfert~2 galaxies have similar SED's in the near-IR to radio
wavelengths.  However, they have a large spread in the UV range, being
as red as a normal galaxy or as blue as a Starburst. This increasing
blueness can be due to an increasing contribution from the AGN
continuum to the spectrum, like in NGC1068, or to the presence of
circumnuclear HII regions, like NGC7130.  Figures 2 and 3 also show
that there is a steep drop in the emission from far-IR to the
millimeter waveband (Log$\nu\approx$11.5). This drop, which is similar
to the one observed in quasars (Sanders et al. 1989; Antonucci et al.
1990), represents the end of the thermal emission from radiation
reprocessed by the circumnuclear torus and maybe HII regions in the
galaxy disk, and the beginning of the non-thermal, synchrotron radio
emission.

The LINER SED's are similar in the radio and visual part of the
spectrum, but have some spread in mid$/$far IR and UV wavebands.  The
mid$/$far IR spread can be explained using the same arguments used
above for normal galaxies, while the difference in the UV band can be
due to an increasing contribution from a population of young stars, or
the active nucleus.  The emission in the X-ray has some spread due to
the large aperture.

The SBL's and SBH's have similar SED's along the entire energy
spectrum.  The SBL's have a small spread in the UV, while for SBH's the
most noticeable spread is in the radio and far IR bands. The X-ray
emission, contrary to what is observed for the rest of the galaxies,
drops abruptly relative to the UV emission in both types of Starburst
galaxies.  The emission in the X-ray comes mostly from SNR,
concentrated in the Starburst region.

\section{Comparison Between Different SED's}

In Figure 4 we make a comparison between objects of similar activity
class, normalized again at $\lambda$7000\AA.  On the bottom panel we
compare the average SED of normal Ellipticals and Spirals. The two
groups are very similar from the radio to the visual waveband. The most
significant difference is in the ultraviolet part of the spectrum,
where the Spirals have an increasing contribution from HII regions. The
apparent large difference at Log$\nu\approx$13.5 may be due to the fact
that at this waveband, flux was available only for some of the
Ellipticals and for no Spirals. Likewise, the difference in the X-ray
fluxes is uncertain due to the small number of Spirals with available
X-ray fluxes.

On the middle panel we compare the SED's of LINER's and Seyfert 2's.
The two SED's overlap only in the visual to near-IR region
(14$<$Log$\nu<$15), where they are dominated by the old stellar
population, differing in all other wavebands (but see below, where
LINER's and Seyfert 2's are compared using different normalizations).
The UV and X-ray emission of Seyfert 2's is larger than that of
LINER's, consistent with a larger contribution from the active nucleus,
or in some cases, the presence of a circumnuclear HII region. The
Seyfert 2's are also brighter than LINER's in the mid$/$far IR and
radio wavebands. Most of the IR emission in Seyfert 2's is probably due
to reradiation of the nuclear emission by a circumnuclear torus
(Storchi-Bergmann, Mulchaey \& Wilson 1992), which is possibly not
present in LINER's.  The higher radio emission from the Seyfert 2's can
be explained by the higher nuclear activity of these objects.

On the top panel of Figure 4 we compare the SED's of SBH's and SBL's.
These two SED's are very similar along the entire energy spectrum. The
only differences are in the ultraviolet, where the SBL's are brighter
than SBH's due to the lower reddening, and in the mid$/$far IR, where
SBH's are brighter than SBL's. This behaviour was studied by Calzetti
et al. (1995), who found that the energy absorbed in the UV is
reradiated in the mid$/$far IR.

In Figure 5 we show the comparison among groups of different activity
class.  On the top left panel we plot the Seyfert 2's, SBL's and SBH's
SED's. These SED's are similar from the radio to near-IR waveband.
However, they start to diverge in the visual towards UV wavelengths. In
this waveband the Seyfert 2's are dominated by the old stellar
population and have the reddest energy distribution, probably due to
the obscuration of the AGN continuum by the torus, while SBH's and
SBL's are increasingly bluer, and dominated by the young stellar
population. These SED's also differ in the X-ray waveband where the
Seyfert 2's are brighter.  On the top right panel we show the LINER's,
SBH's and SBL's SED's. The only wavelength region where these SED's are
similar is from the visual to the near-IR, where they are normalized.
The LINER's SED is systematically fainter at all other bands.

On the bottom left panel of Figure 5 we show the SED's of LINER's,
Seyfert 2's and Spirals. The LINER's and Spirals have similar SED's,
only differing in the mid$/$far IR and UV, where the Spirals are
fainter than the LINER's. Seyfert 2's and Spirals SED's are similar
only in the near-IR to visual waveband, where they are dominated by the
old stellar population. The Seyfert 2's are much brighter than the
Spirals in the IR and UV.  The SED's of Spirals, SBL's and SBH's are
compared on the bottom right panel of Figure 5. Here we can see the
difference between SED's dominated by old (Spirals) and young stellar
populations (SBH's and SBL's). The only wavelength region where these
SED's can be considered similar is in the visual to near-IR, again the
region where they are normalized.  In these region the Starbursts have
some contribution from old stars. The Spirals are fainter in any other
waveband.

In Figure 6 we compare the SED's of LINER's and Seyfert 2's (top),
SBL's and SBH's (bottom) with those of Radio Quiet and Radio Loud
Quasars from Sanders et al. (1989) (RQQ and RLQ hereafter).  In
contrast to the previous analysis, here the SED's were normalized to
the 60$\mu$m flux.  We chose $\lambda$60$\mu$m as normalization
wavelength because this is the wavelength region that is the most
isotropic in the entire quasars SED (Pier \& Krolik 1992).  We could
not find an average SED for Seyfert 1's, but a comparison between the
RQQ SED with that of the Seyfert 1 galaxy NGC3783 (Alloin et al. 1995),
showed that they are very similar.

The comparison between the SED's of Quasars and the other galaxies
shows that Quasars are $\approx$0.5 dex brighter in the mid$/$far-IR,
$\approx$1 dex brighter in the near-IR, and $\approx$2 to 2.5 dex
brighter in the visual to X-ray region of the spectrum. The only
exception to the above differences are for LINER's in the visual to
near-IR region of the spectrum, whose SED's touch those of the Quasars.
This is due to the fact that the nuclear luminosity of LINER's, i.e.
the energy emitted from the nuclear engine is much smaller than that of
Quasars. When the SED's are normalized to the radiation that is emitted
isotropically from the nucleus (60$\mu$m), the near-IR and visual
regions of the SED's of LINER's, which are dominated by the stellar
population in these objects, will be shifted to values comparable to
those of Quasars.  In the radio waveband, the RQQ's SED is similar to
that of Seyfert 2's, Starbursts and LINER's, while the RLQ's SED's are
$\approx$3 dex brighter than all others.
From Figure 6 we see that the RQQ and RLQ SED's are dominated by the
visual
and UV emission, which is due to the nuclear featureless continuum. As
opposed to the Seyfert 2's, LINER's and Starbursts, Quasars do not have
a pronounced mid$/$far IR emission bump relative to the visual and UV
parts of the spectrum.

Another interesting fact to be noticed in this Figure is the similarity
between the SED's of Seyfert 2's and LINER's, when we normalize them to
the 60$\mu$m flux. With the exception of the visual and near-IR region
of the spectrum where, due to their low nuclear luminosity LINER's are
dominated by the stellar population, the two SED's are very similar,
suggesting that LINER's are indeed low luminosity relatives of
Seyferts.

\section{The SED of HII Regions and Supernova Remnants}

Here we describe the SED of HII regions, a thermal and a non-thermal
Supernova Remnant (SNR). These SED's can be compared with those from
Starbursts, in order to determine the wavebands where the young
components contribute most to the SED.

As examples of single HII regions we use NGC5455, NGC5461 and NGC5471,
in the disk of M101, and NGC604 in the disk of M33.  These objects are
bright, have sizes of 30\sec\ typically, and are not resolved into
stars, which make them ideal for our analysis. Their metallicities are
subsolar, 12$+$logO$/H=$8.51, 8.28, 8.31 and 8.05 for NGC604, NGC5455,
NGC5461 and NGC5471, respectively (Garnett 1989; Torres-Peimbert,
Peimbert \& Fierro 1989). For the non-thermal SNR we use the Crab
nebula, which is a close and well studied object, while for the thermal
SNR we use N49 in the LMC, which is relatively compact and bright.

The X-ray fluxes of the HII regions, observed with ROSAT, were obtained
from Williams \& Chu (1995) for the objects in M101 and Schulman \&
Bregman (1995) for the objects in M33. The UV fluxes, observed with
IUE, were measured from Figures 6, 29, 30 and 32 of Rosa, Joubert \&
Benvenutti (1984) for NGC604, NGC5455, NGC5461 and NGC5471,
respectively. The radio fluxes at 1.47 GHz and 4.89 GHz were obtained
from Sramek \& Weedman (1986), and are integrated over the entire HII
region. The mid$/$far-IR (IRAS) fluxes were obtained from NED.  The
near-IR fluxes (J,H and K) of NGC5455 and NGC5471 were obtained from
Campbell \& Terlevich (1984), observed with an aperture of 10\sec.  For
NGC5461 we use the values from Blitz et al. (1981), obtained with an
aperture of 10\sec, while for NGC604 we use the values from Hunter \&
Gallagher (1985), observed with an aperture of 23\sec.

The visual fluxes of NGC604 were measured from Figure 6 of D'Odorico,
Rosa \& Wampler (1983). They observed several parts of the HII region,
with apertures of 4\sec$\times$8\sec, and give the sum of these
observations, which corresponds to an aperture similar to that of IUE.
For NGC5455, NGC5461 and NGC5471, the visual fluxes were calculated
from Torres-Peimbert et al. (1989), using their emission-line fluxes
and equivalent widths. Their aperture was 3.8\sec$\times$12.4\sec,
which corresponds to $\approx$25\% of the IUE aperture, but include the
HII region peak emission.

The SED of the Crab nebula was obtained from Woltjer (1987) and is
described by the following relations. For 7$<$Log$\nu<$12,
Log$\nu$F$_{\nu}=$5.717$+0.7\times$Log$\nu$; for 13.3$<$Log$\nu<$15.5,
Log$\nu$F$_{\nu}=$13.01$+0.15\times$Log$\nu$; and for 16$<$Log$\nu<$19,
Log$\nu$F$_{\nu}=$17.797$-0.15\times$Log$\nu$.  The flux densities
($\nu$F$_{\nu}$) in the IRAS bands were measured from Figure~4 in that
paper and are: 15.08, 15.15, 15.06 and 14.88 for the wavebands 12, 25,
60 and 100$\mu$m, respectively.

The radio data of N49 were obtained from Wright \& Otrupcek (1990) and
are 2.73 Jy (0.48 GHz), 1.16 Jy (2.7 GHz), 0.63 Jy (5.0 GHz) and 0.47
Jy (8.4 GHz).  The mid$/$far-IR (IRAS) fluxes are 0.56 Jy (12$\mu$m),
1.78 Jy (25$\mu$m), 19.5 Jy (60$\mu$m) and 41.6 Jy (100$\mu$m)
(Schwering \& Israel 1990).

X-ray, UV and visual fluxes of N49 were obtained from Vancura et al.
(1992).  The X-ray flux, observed with Einstein and integrated over the
entire SNR, is 6.34$\times$10$^{-11}$ erg cm$^{-2}$ s$^{-1}$.  The flux
in the visual band, obtained from a narrow-band image centered at
$\lambda$6100\AA\ and corrected for internal reddening (E(B-V)$=$0.35)
using the extinction law of Fitzpatrick (1986), is
3.85$\times$10$^{-14}$ erg cm$^{-2}$ s$^{-1}$ \AA$^{-1}$.  In order to
obtain the UV fluxes for the entire SNR we use the fact that the
$\lambda$6100\AA\ flux inside Vancura et al. (1992) ``A'' IUE aperture
is 10\% that of the entire nebula, and assume that this percentage is
equal for the UV waveband. The UV fluxes of the ``A'' aperture were
measured from their Figure~6, multiplied by 10, and corrected for
internal reddening. The final fluxes are 4.63$\times$10$^{-12}$,
1.1$\times$10$^{-12}$ and 4.87$\times$10$^{-13}$ erg cm$^{-2}$ s$^{-1}$
\AA$^{-1}$ for 1350\AA, 2200\AA\ and 2900\AA, respectively.

The fluxes of individual HII regions, as well as the average SED of HII
regions, N49 and Crab nebula are given in Table 7. We show in Figure 7
the individual HII regions SED's normalized to the flux at
$\lambda$7000\AA. These SED's are very similar along the entire energy
spectrum, showing a steep ultraviolet continuum, a small bump in the
near-IR (Log$\nu\approx$14 Hz) and a large bump in the mid$/$far IR.
However, the near-IR bump is uncertain, due to the different apertures
through which the visual and near-IR data were obtained. This same
problem may be affecting the mid$/$far IR bump, since the IRAS
apertures are much larger than the HII regions and can include emission
from warm and cold dust in the galaxy disk (notice that NGC5455 do not
have IRAS data available).

In Figure 8 we compare the SED's of the thermal SNR, non-thermal SNR
and average HII regions, normalized to the HII regions flux at radio
6cm.  The non-thermal SNR has a flat SED from the X-rays to the
infrared waveband.  It has some thermal emission in the mid IR (Woltjer
1987) and drops towards the radio waveband. The thermal SNR has a steep
UV to optical SED, like the HII regions.  This emission comes from H
and He recombination radiation and two photons continuum emission
(Vancura et al. 1992).  The flux drops from UV to X-ray, where it is
similar to that of non-thermal SNR and stronger than HII regions.  The
thermal SNR SED also shows an increase in the far-IR emission due to
cold dust reradiation, and then drops to the radio waveband.

In Figure 9 we compare the SED's of SBL's and SBH's with those of HII
regions (left panel) and SNR's (right panel). The HII regions and
Starbursts are normalized relative to the 25$\mu$m flux, instead of the
$\lambda$7000\AA, because the 25$\mu$m corresponds to the warm dust
emission, which should be similar in these two classes of objects. The
SNR's were again normalized to the flux of HII regions at radio 6cm
(log$\nu=$9.7).

HII regions and Starbursts have similar SED's in the mid$/$far IR parts
of the spectrum, but differ in the X-ray and radio, where HII regions
have smaller fluxes.  In the UV part of the spectrum HII regions are
similar to SBL's, but are much brighter than SBH's. However, Starbursts
are stronger than HII regions in the visual and near-IR parts of the
spectrum.  This difference is due to the fact that in HII regions we
observe only the young stellar population, while in Starbursts we
observe a significant amount of underlying old stellar population,
which contributes mostly to the visual and near-IR parts of the SED.

The comparison between the SED's of SNR's and Starbursts shows that
non-thermal SNR's and Starbursts are similar only in the radio, with
the non-thermal SNR being stronger in X-rays and fainter in the other
wavebands. The thermal SNR and the Starbursts have similar SED's in the
radio.  Thermal SNR's are fainter than Starbursts in the visual to
mid$/$far-IR, but fainter in the UV and X-rays.

\section{Are the different activity classes distinguishable by their
SED's?}

Can we distinguish between different classes of galaxies based on their
SED's? In order to statistically study this we have chosen several
wavebands, normalized to the $\lambda$7000\AA\ flux, and compared the
different SED's using Student's t-test. We use hare the normalization
to the $\lambda$7000\AA\ flux because it represents a normalization to
the old stellar population. However, it should be kept in mind that a
different normalization would produce different results, such as with
the normalization of LINER's and Seyfert 2's at 60$\mu$m.  Table 8
shows the wavebands used and the number of galaxies with those
wavebands available in each group.  The results of the comparison are
shown in Table 9 and Figure 10, where we give the probability of two
SED's being equal.  Two SED's are considered to be significantly
different when the t-test gives probabilities smaller than 0.05 (5\%),
which corresponds to 2$\sigma$ difference. This value is noted with a
line in Figure 10. If the probability is between 0.05 and 0.2 (between
$\approx$1.3 and 2.0 $\sigma$), the SED's are considered to be
moderately different, which means that this difference can be
considered as a tendency, but should be used with caution to
distinguish between two different activity classes.  Notice that we are
not comparing the 6 cm and X-ray emission of normal Spirals with other
galaxies, because there is only a small number of Spirals detected in
these wavebands.

On the bottom panel of Figure 10 we compare objects of similar activity
class. In agreement with the results of the previous section, SBH's and
SBL's can be well distinguished in the visual UV and far-IR parts of
the spectrum.  Seyfert 2
and LINER SED's are significantly different only at 25$\mu$m. However,
with the exception of the near-IR optical band (14$<$log$\nu<$14.5 Hz),
where they are very similar, the probability of the two SED's being
equal, in the remaining wavebands the difference is only moderately
significant.  The comparison between Ellipticals and Spirals shows that
their SED's are very similar. Only in the UV (1355\AA) the probability
of the two distributions being equal reaches values smaller than 0.15.

On the middle panel we compare Active (Seyfert 2, LINER, SBH and SBL)
with normal Spiral galaxies SED's. We chose to compare the active
galaxies only to the normal Spirals, because the Ellipticals SED's are
very similar to them, and also because the host galaxies of the Active
objects are spirals. The Spirals can be separated from SBH's and SBL's
in the mid$/$far IR, visual and UV wavebands. The comparison with the
Seyfert 2 template shows that the two SED's can be well separated in
the mid$/$far IR and also in the UV (2900\AA). LINER's and Spirals are
similar along most of the energy spectrum. Only in the mid$/$far IR is
the probability of the two distributions close enough to 0.05 for them
to be considered as moderately different.

On the top panel of Figure 10 we compare the SED's of SBH's and SBL's
with Seyfert 2's and LINER's. Seyfert 2's SED is different from both
SBH's and SBL's in the visual and UV waveband, and also different from
SBL's in the near-IR. It can be considered as moderately different from
SBH's in the X-rays and near-IR.  The LINER's SED is different, or
moderately different from that of SBL's in the UV to mid IR range. When
compared to SBH's, LINER's are different in the UV, visual and
mid$/$far IR wavebands.

In conclusion, the statistical analysis confirms the qualitative
results from the previous sections.  The largest differences over the
entire 10 decades of frequency exist between LINER's and SBL's. In all
other cases, the differences are limited to specific ranges, such as
those between Seyfert 2's and LINER's in the mid$/$far IR and UV.
Normal galaxies can be separated from active ones (Starbursts, LINER's
and Seyfert 2's) by the lower mid$/$far IR, and UV emission, relative
to the visual. Seyfert 2's and LINER's can be easily differentiated
from Starbursts, based on their smaller UV$/$visual ratio.

\section{Bolometric Fluxes}

The bolometric fluxes were calculated by integrating the SED's.  The
contribution of the X-ray band to the bolometric luminosity is very
small, and consequently does not affect the results for those galaxies
without data available in this waveband. A comparison between the
bolometric fluxes and galaxy diameters shows that these quantities are
independent. This result assures us that the flux of wavebands like the
mid$/$far IR, which were observed through apertures much larger than
that of the IUE, are not shifting the bolometric flux of large objects
to higher values.

In Figure 11 we compare the bolometric flux with the flux density of
selected wavebands. Considering all galaxies together, the 100$\mu$m
flux density  shows the best correlation with the bolometric flux.
When we consider only galaxies of the same activity class, their
bolometric fluxes also show a good correlation with the flux density in
other wavebands.  We can also see in this Figure that the wavebands
which contribute most to the bolometric flux in Seyfert 2's, LINER's
and Starbursts are the mid$/$far IR.  For normal galaxies, the emission
from these wavebands is weaker and the wavebands which contribute most
to the bolometric flux are the near-IR and visual.

The observed correlation can be used to obtain the bolometric flux of
galaxies with different activity classes, based on information of a
limited wavelength range. In order to quantify this, we separate the
galaxies in groups, according to activity class:  Normal galaxies
(Spirals $+$ Ellipticals), Seyfert 2's, SBL's and SBH's.  LINER's are
excluded from this analysis because of the small number of objects in
the sample.  For these groups we perform linear fits of the form
$Log(F_{bol})=a+b\times Log(\nu F_{\nu})$.

The resulting coefficients ``a'' and ``b'', as well as the correlation
coefficients of the linear fits are given in Table 10.  For normal
galaxies, the near-IR wavebands are the ones which better correlate
with the bolometric flux. For Seyfert 2's the bolometric flux
correlates well with the flux in the mid$/$far IR bands.  SBL's
bolometric flux correlates well with the fluxes of the wavebands in the
range 2530\AA\ to far-IR, while for SBH's the best correlation is in
near and far-IR.

\section{Summary}

In this paper we built the radio to X-ray SED's of 59 galaxies,
including normal Spirals, Ellipticals, LINER's, Seyfert 2's and
Starbursts.  Also, for the comparison with Starbursts, we built SED's
for HII regions, thermal and non-thermal SNR's. We used data selected
from the literature, trying to match the IUE aperture
(10\arcsec$\times$20\arcsec), and discuss the possible contamination
effects for the wavebands observed with larger apertures.

The SED's were normalized to the flux at $\lambda$7000\AA, which
corresponds to a normalization by the old stellar population, and
averaged according to their activity and morphological classes.  Both a
qualitative and a quantitative comparison between the SED's of
different classes of objects were performed, giving similar results,
which can be summarized as follows. The normal Spirals and Ellipticals
have similar SED's over the entire energy range, but are fainter than
the other SED's, relative to the $\lambda$7000\AA\ flux. The Seyfert 2
SED's are similar to those of LINER's in the visual and near-IR, but
stronger in the other wavebands. When compared to Starbursts, Seyfert
2's have similar SED's in the radio to near-IR, are weaker in the
ultraviolet, but stronger in the X-rays. The SBH's and SBL's SED's are
very similar along the entire energy range, with the exception of the
ultraviolet, where SBH's are weaker, and mid$/$far IR, where they are
stronger. These differences can be accounted to the higher absorption
and reradiation of the ionizing radiation in SBH's.

The SED's of Seyfert 2's, LINER's and Starbursts were compared with
SED's of RQQ and RLQ, normalized to the flux at $\lambda$60$\mu$m. The
Quasars SED's are between 1 and 2 dex stronger than the other SED's,
depending on the waveband. The exception occurs for RQQ SED's, which
are similar to those of the other galaxies in the radio to far-IR
wavebands.
From this comparison we have also found that, when using the
normalization
at $\lambda$60$\mu$m, the SED's of LINER's and Seyfert 2's are very
similar, with the exception of the optical to near-IR wavebands where
LINER's are dominated by the old stellar population.

We have also constructed SED's of HII regions, thermal and non-thermal
SNR's. HII regions and thermal SNR's have similar SED's and differ only
in the X-rays, where HII regions are fainter, and far-IR, where HII
regions are stronger. The SED of the non-thermal SNR is a flat
continuum, for which we do not have a good normalization point to
compare with the other SED's.  The comparison of Starbursts with HII
regions shows that they are very similar, with the exception of the
X-rays, visual and near-IR, where Starbursts are stronger, due to the
contribution from old stars in the visual and near-IR, and
``superwinds'' in X-rays (Heckman, Armus \& Miley 1990).

Finally, we calculated the bolometric fluxes of the galaxies and
compared them with the flux densities of individual wavebands. From
this comparison we found that the mid$/$far IR emission dominates the
energy output in Seyfert 2's, LINER's and Starbursts. For Spirals and
Ellipticals the visual and near-IR emission contributes most to the
bolometric flux.  We have also performed linear regressions between the
bolometric fluxes and flux densities, which can be used to determine
the bolometric flux of objects with reduced waveband information.

\acknowledgements

This work was supported by NASA under grant NAGW-3757 and by the
Brazilian institution CNPq. This research has made use of the
NASA$/$IPAC Extragalactic Database (NED) which is operated by the Jet
Propulsion Lab, Caltech, under contract with NASA. We would like to
thank N. Panagia and K. Long for useful discussions about supernova
remnants.

\clearpage

\begin{figure} \caption{Individual SED's, separated by arbitrary
constants.  The galaxy name is shown on the left of each SED. The
dashed lines represent regions for which there were no Iras data
available. a) Normal Ellipticals (top) and Spirals (bottom); b)
Seyfert~2's (top) and LINER's (bottom); c) Low Reddening Starburst's
(top) and High Reddening Starbursts (bottom).} \end{figure}

\begin{figure} \caption{Plot of individual SED's, normalized to the
flux at 7000\AA, of Normal Ellipticals (top left), Normal Spirals (top
right), Seyfert 2's (middle left), LINER's (middle right), Low
reddening Starbursts (bottom left) and High Reddening Starbursts
(bottom right).} \end{figure}

\begin{figure} \caption{Plot of the average SED's, using the same order
as Figure 2.  The error bars are the standard deviation of the
average.} \end{figure}

\begin{figure} \caption{Comparison between the average SED of Normal
Ellipticals and Spirals (bottom); Seyfert~2's and LINER's (middle); and
High and Low Reddening Starbursts (top).} \end{figure}

\begin{figure} \caption{Comparion between the average SED of Seyfert~2,
High and Low Reddening Starbursts (top left); LINER's, High and Low
Reddening Starbursts (top right); Normal Spirals, LINER's and
Seyfert~2's (bottom left); and Normal Spirals, High and Low Reddening
Starbursts (bottom right).} \end{figure}

\begin{figure} \caption{Comparison of the SED of Seyfert~2's and
LINER's with Radio Quiet and Radio Loud Quasars  from Sanders et al.
(1989), normalized to the flux at $\lambda$60$\mu$m (top); High and Low
Reddening Starbursts with Radio Quiet and Radio Loud Quasars (bottom).}
\end{figure}

\begin{figure} \caption{SED's of single HII regions, normalized to the
flux at $\lambda$7000\AA.} \end{figure}

\begin{figure} \caption{SED's of HII Regions, a Non-Thermal SNR (Crab
Nebula) and a Thermal SNR (N49 in the LMC), normalized to the flux at
$\lambda$7000\AA.} \end{figure}

\begin{figure} \caption{The comparison of SBL and SBH SED's with the
SED's of HII regions (left),normalized to the flux at
$\lambda$25$\mu$m, thermal and non-thermal SNR's (right), normalized to
the flux at $\lambda$7000\AA.} \end{figure}

\begin{figure} \caption{Probability of two SED's being equal as a
function of the waveband.  The horizontal line at 0.05 represents the
probability below which two SED's can be considered different. When the
probability is between 0.05 and 0.2 the SED's are moderately
different.} \end{figure}

\begin{figure} \caption{Relations between Bolometric flux and the flux
densities at six wavebands, 100$\mu$m (left bottom), 25$\mu$m (left
middle), 2.2$\mu$m (left top), 1.6$\mu$m (right bottom), 1.2$\mu$m
(right middle) and 7000\AA\ (right top). The vertical axis has units of
ergs cm$^{-2}$ s$^{-1}$.  Filled squares represent Seyfert 2's, open
squares LINER's, filled triangles normal Ellipticals, open triangles
normal Spirals, filled circles SBH's and open circles represent SBL's.}
\end{figure}

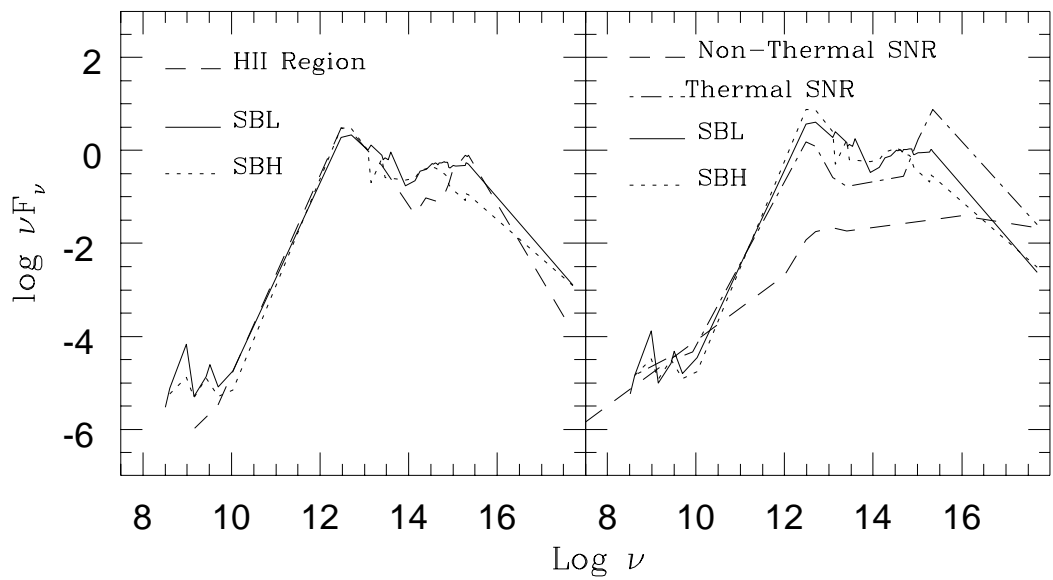
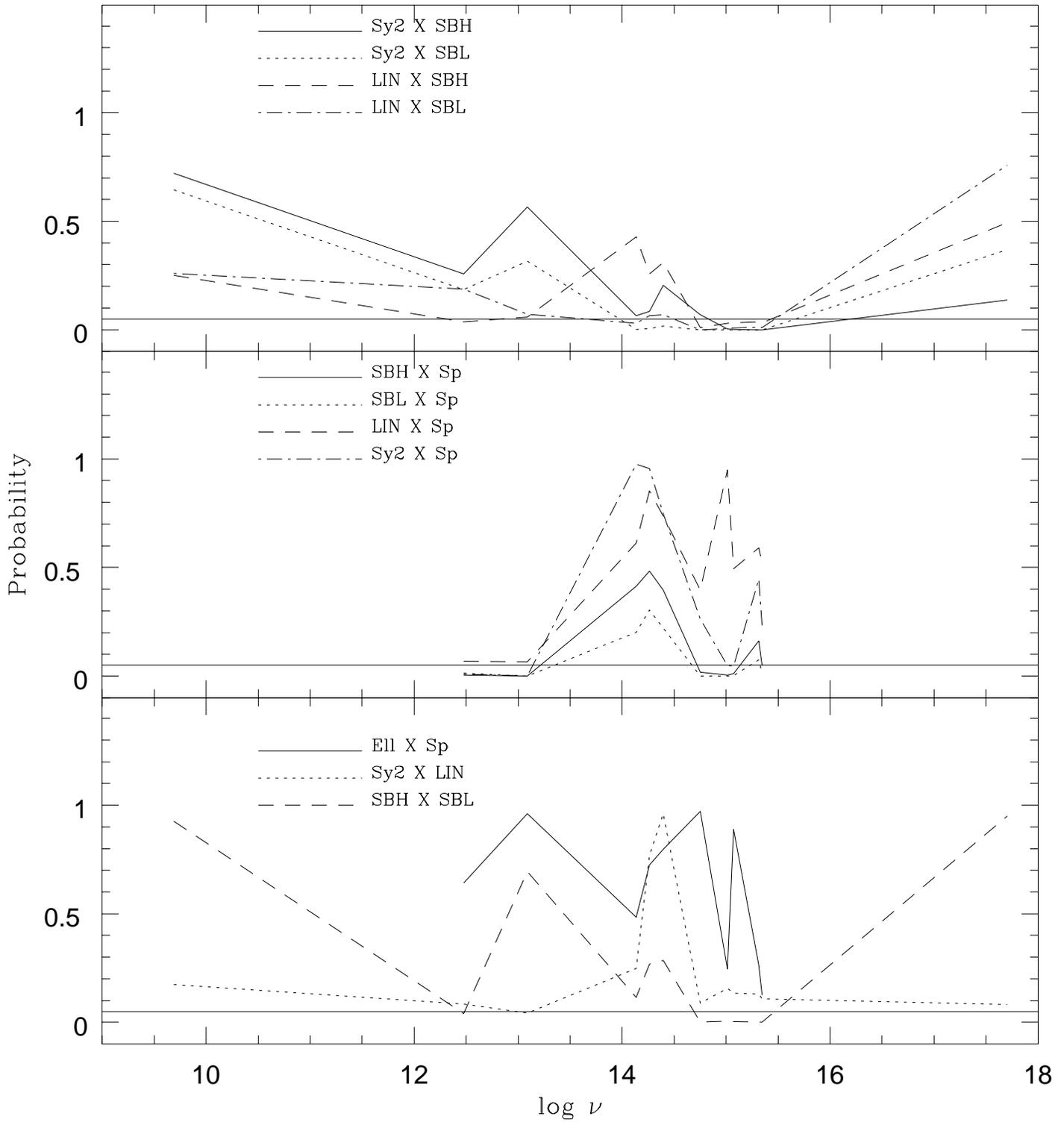
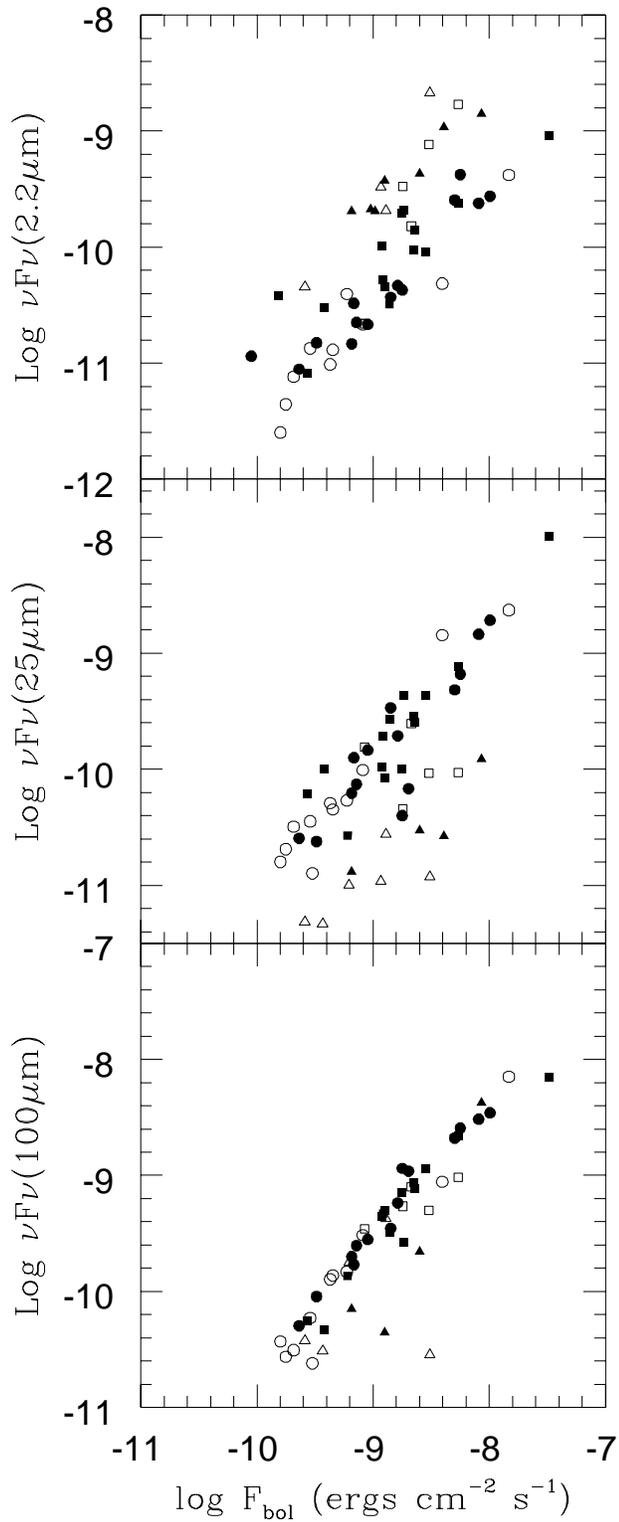
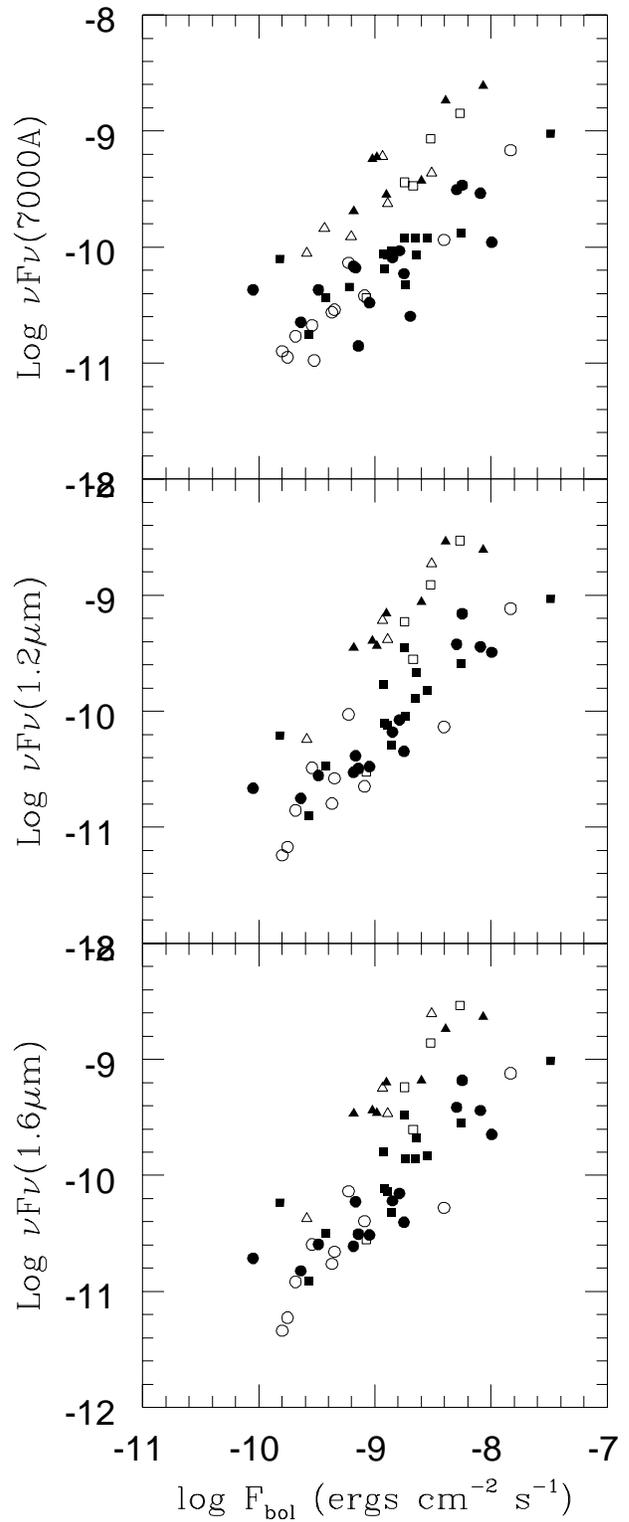
\end{document}